\documentclass[sigconf]{acmart}

\usepackage{algorithmic}
\usepackage{graphicx}
\usepackage{textcomp}
\usepackage{xcolor}
\usepackage{siunitx}
\usepackage{hyperref}
\usepackage[nolist]{acronym}
\usepackage{placeins}

\usepackage{caption}
\usepackage{subcaption}

\DeclareSIUnit\dbi{dBi}                                             
\DeclareSIUnit\dbm{dBm}                                             
\DeclareSIUnit\Ah{Ah}                                               
\newcommand*{\rom}[1]{\expandafter\@slowromancap\romannumeral #1@}  

\AtBeginDocument{%
  }

\copyrightyear{2023} 
\acmYear{2023} 
\setcopyright{acmlicensed}\acmConference[ENSsys' 23]{11th International Workshop on Energy Harvesting \& Energy-Neutral Sensing Systems}{November 12, 2023}{Istanbul, Turkiye}
\acmBooktitle{11th International Workshop on Energy Harvesting \& Energy-Neutral Sensing Systems (ENSsys' 23), November 12, 2023, Istanbul, Turkiye}
\acmPrice{15.00}
\acmDOI{10.1145/3628353.3628549}
\acmISBN{979-8-4007-0438-3/23/11}

\acmConference[ENSsys' 23]{11th International Workshop on Energy Harvesting \& Energy-Neutral Sensing Systems}{November 12, 2023}{Istanbul, Turkiye}

\usepackage{eso-pic}
\usepackage{url}
\AddToShipoutPictureBG*{
  \AtPageUpperLeft{%
    \put(0,-40){\raisebox{15pt}{\makebox[\paperwidth]{\begin{minipage}{21cm}\centering
      \textcolor{gray}{This article has been accepted for publication in the proceedings of the \\
      11th International Worksjop on Energy Harvesting and Energy-Neutral Sensys Systems (ENSSys'23).\\ 
      DOI: 10.1145/3628353.3628549} 
    \end{minipage}}}}%
  }
  \AtPageLowerLeft{%
    \raisebox{25pt}{\makebox[\paperwidth]{\begin{minipage}{21cm}\centering
      \textcolor{gray}{ \copyright 2023  Authors and ACM. 
       This is the author’s version of the work. It is posted here for your personal use. Not for redistribution. The definitive Version of Record will
        be published in Proceedings of the 11th International Workshop on Energy Harvesting and Energy-Neutral Sensing Systems, https://doi.org/10.1145/3628353.3628549
      }
    \end{minipage}}}%
  }
}

\begin{document}

\title{A Lora-Based and Maintenance-Free Cattle Monitoring System for Alpine Pastures and Remote Locations}


\author{Lukas Schulthess}
\affiliation{%
  \institution{Dept. of Information Technology and Electrical Engineering, ETH Z\"{u}rich}
  \city{Z\"{u}rich}
  \country{Switzerland}
}
\email{lukas.schulthess@pbl.ee.ethz.ch}

\author{Fabrice Longchamp}
\affiliation{%
  \institution{Dept. of Information Technology and Electrical Engineering, ETH Z\"{u}rich}
  \city{Z\"{u}rich}
  \country{Switzerland}
}
\email{fabricel@ethz.ch}

\author{Christian Vogt}
\affiliation{%
  \institution{Dept. of Information Technology and Electrical Engineering, ETH Z\"{u}rich}
  \city{Z\"{u}rich}
  \country{Switzerland}
}
\email{christian.vogt@pbl.ee.ethz.ch}

\author{Michele Magno}
\affiliation{%
  \institution{Dept. of Information Technology and Electrical Engineering, ETH Z\"{u}rich}
  \city{Z\"{u}rich}
  \country{Switzerland}
}
\email{michele.magno@pbl.ee.ethz.ch}

\begin{abstract}


The advent of the \ac{IoT} is boosting the proliferation of sensors and smart devices in industry and daily life. Continuous monitoring \ac{IoT} systems are also finding application in agriculture, particularly in the realm of smart farming. 
The adoption of wearable sensors to record the activity of livestock has garnered increasing interest. 
Such a device enables farmers to locate, monitor, and constantly assess the health status of their cattle more efficiently and effectively, even in challenging terrain and remote locations.
This work presents a maintenance-free and robust smart sensing system that is capable of tracking cattle in remote locations and collecting activity parameters, such as the individual's grazing- and resting time.
To support the paradigm of smart farming, the cattle tracker is capable of monitoring the cow's activity by analyzing data from an accelerometer, magnetometer, temperature sensor, and \ac{GNSS} module, providing them over \ac{LoRaWAN} to a backend server. 
By consuming \qty{511.9}{\joule} per day with all subsystems enabled and a data transmission every 15 minutes, the custom-designed sensor node achieves a battery lifetime of 4 months. 
When exploiting the integrated solar energy harvesting subsystem, this can be even increased by 40\% to up to 6 months.
The final sensing system’s robust operation is proven in a trial run with two cows on a pasture for over three days.
Evaluations of the experimental results clearly show behavior patterns, which confirms the practicability of the proposed solution.

\end{abstract}




\maketitle

\section{Introduction}\label{sec:introduction}
The \ac{IoT} with its vastly deployed sensors and smart devices has long since found its way into the industry and our daily lives.
Sensors continuously collecting data help to improve efficiency, increase throughput, reduce expenses, and prevent structural failures in fabrication processes.
Home automation systems improve living comfort, provide security, and increase the quality of life with the help of smart sensor systems.  
In asset tracking and goods monitoring, wireless data loggers are used in goods production lines \cite{horejsi_smart_factory_2020}, delivery services \cite{eeshwaroju_drone_delivery_2020, sayeed_drig_delivery_2019}, controlling-, and monitoring systems \cite{nuzzo_structural_health_2021}. When sensor systems are used extensively, it is possible to gather enormous amounts of data from the environment. Data analysts utilize such sensor systems to collect data generated in urban environments to improve traffic flows, reduce energy consumption, and assess the utilization and attractiveness of public spaces, allowing for optimal resource management \cite{schulthess_smartcity_2023}.

Whereas sensor systems have been well-established and are widely used in the industry, planning- and social sectors, the agricultural sector also benefits from such continuous monitoring systems \cite{alzubi_smart_farming_2023, reddy_smart_farming_2021, farooq_iot_agriculture_2020}. With Smart Farming - the concept of using advanced technology in agriculture to monitor, track, and automate operations - resource management can be optimized and simplified. This is especially important for livestock control \cite{joshitha_lifestock_control_2021}.
Having wearable sensors to record animal activity in herds has shown increasing interest over the last few years.
Automated livestock monitoring using activity trackers simplifies cattle monitoring and can provide relevant information about the individual's activity \cite{tran_animal_behaviour_recognition_2022, cabezas_cattle_behaviour_2022, rahman_cattle_behaviour_2017} and health status in a herd \cite{singhal_cattle_collar_2022, jose_animal_welfare_2020}. Injuries, infections, and diseases can be detected and treated early \cite{porto_cattle_diseases_detect_2021}.
Not only the activity parameters are interesting for farmers, but knowing the location of every individual helps hold the herd together, identify their owner, and even prevent cattle theft \cite{jegan_cattle_tracking_2022, facina_cattle_theft_2022, veintimilla_cattle_position_2022}

Having wearable sensors to monitor animal activity and well-being is especially attractive where there is only sporadic contact between the farmer and the herd's individuals \cite{zhumani_cattle_tracker_2022}. In alpine regions, farmers face a significant challenge regarding cow monitoring; the rugged and steep landscape characterizes the majority of the agricultural landscape and limits access to alpine pastures. In addition, such alpine meadows are usually very spacious and - due to the terrain - not always possible to fence. This raises the chances of a cow harming itself and, when combined with intermittent observation, raises the risk of life-threatening accidents. To address this challenge, farmers are in need of tracking systems that allow them to remotely locate their cattle.

The ability to perform reliably over long periods is an important criterion for such cattle trackers.
Usually, the primary component of maintenance work for sensor nodes is the replacement or recharging of batteries \cite{ma_eh_iot_2020}. For cattle trackers, maintenance during the deployment is no option, as it would conflict with the tracker's fundamental use case. 
Thus, it is essential to consider energy efficiency when designing such a tracking device, including optimizations on both,
software- and hardware levels. 
Implementing energy harvesting capabilities, for example from solar cells, further relaxes the power budget \cite{srbinovski2015energy}.
A carefully selected communication technology, in combination with well-balanced update rates, further reduces the sensor nodes’ power consumption during data transmission \cite{magno_wulora_2017, mayer2020smart}. 
Finally, animal trackers must be durable and able to resist environmental conditions during their deployment. 
While studies frequently take place in controlled settings, this may not accurately represent how systems behave in real application scenarios. As a consequence, an on-site examination is essential.

\begin{figure}[t!]
    \centering
    \includegraphics[width=\columnwidth]{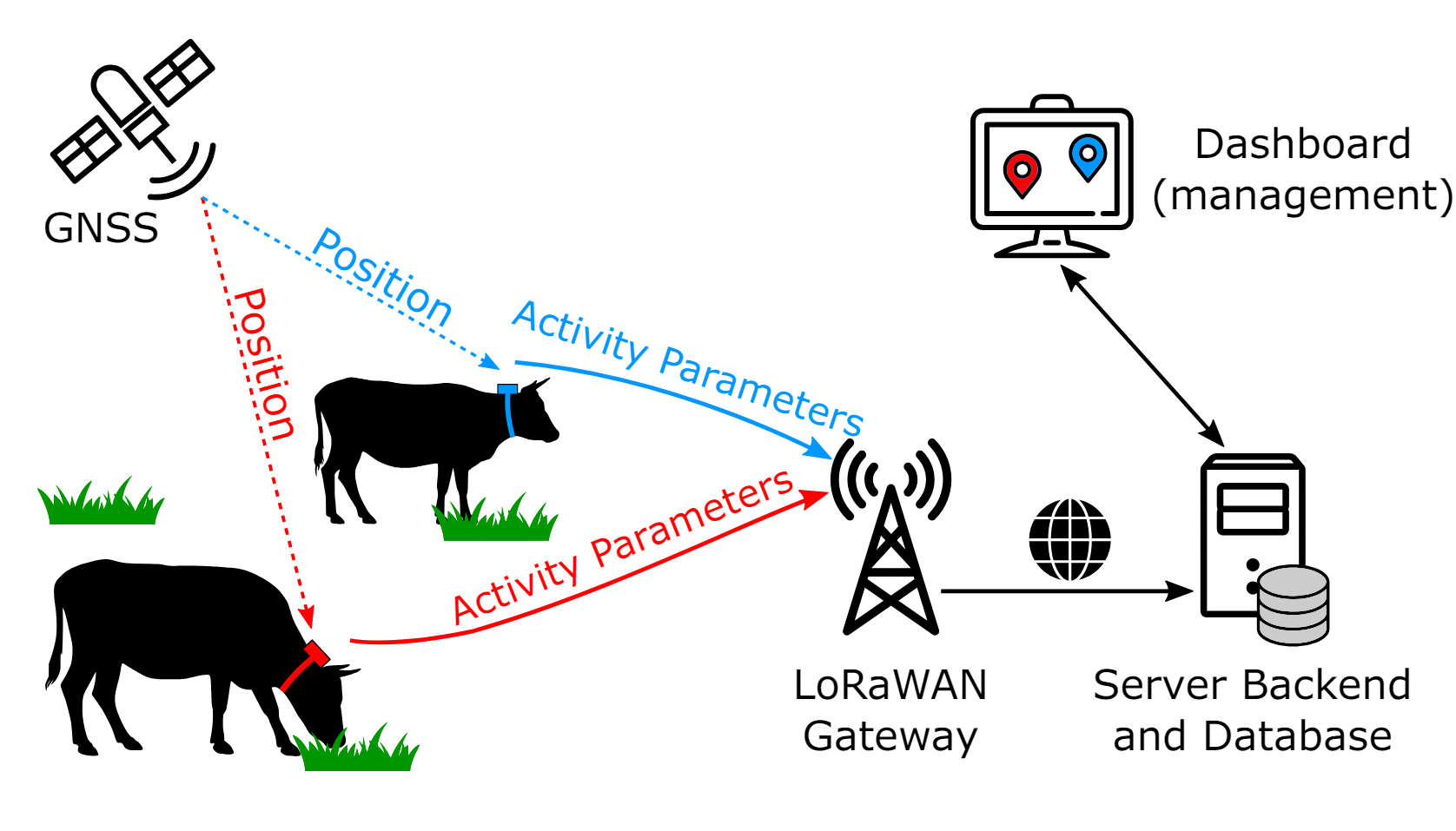}
    \vspace{-10mm}
    \caption{System-wide overview of multiple cattle trackers connected to the server back-end over a LoRaWAN gateway. The collected activity parameters of each active device are stored in an SQL database.}
    \label{fig:system_overview}
\end{figure}

Within this context, this paper presents the following:
\begin{itemize}
    \item Custom end-to-end implementation, and evaluation of a \ac{LoRa}-based sensing system for cattle tracking and monitoring on alpine pastures supplied by rechargeable batteries and solar energy harvesting enabling a maintenance-free operation for one alpine period (5 months at the maximum, starting end of April and ending in October \cite{alpine_season_2022})
    \item Environmental data collection and on-board processing capabilities to determine the cow's activity, such as grazing time and distance traveled, including a back-end for data recording and additional statistical evaluation. 
    \item Successful deployment and test by collecting activity parameters data such as head movements, \ac{GNSS} position, and environmental temperature over a time period of three days from two cows. 
    \item ruggedized and water-proof housing for reliable functionality, in order to accommodate the fast-changing weather conditions in the alpine regions
\end{itemize}

The subsequent paper is organized as follows; existing works on animal trackers are summarized in \autoref{sec:relatedwork}.
The hardware- and software architecture of the proposed cattle tracking system is described in \autoref{sec:system_adefinition}. \autoref{sec:system_evaluation} presents the evaluation of the system in a real application scenario, Finally, the conclusion is drawn in \autoref{sec:conclusion}.

\section{Related Work}\label{sec:relatedwork}

Animal-borne data loggers and location trackers show an increase in interest for both wildlife biologists and livestock researchers.
Independent of their specific use case, starting from small animals like birds held in a controlled environment \cite{schulthess_tinybird_2023}, over mid-sized animals like sheep on large pastures \cite{nobrega_2018}, up to large lions living in the wild \cite{wijers_lions_2018}, they share the common goal of collecting data to get valuable insights into the individual's behavior, health, group dynamics, and fitness. 

In the context of smart farming, livestock trackers mainly focus on the goal of tracking livestock's individual locations and monitoring their health.
Most devices in literature use a combination of accelerometer, magnetometer, gyroscope, and \ac{GNSS} module to acquire data and process them either directly on the tracker itself using a tiny \ac{ML} algorithm \cite{cabezas_cattle_behaviour_2022} or in the cloud \cite{singhal_cattle_collar_2022}. 
For cattle, possible positions for placing such sensors are ears using small and lightweight ear tags \cite{munoz_cow_ear_tag_2018, tsenkov_cow_ear_tag_2017}, the neck using a collar \cite{, veintimilla_cattle_position_2022, facina_cattle_theft_2022, pratama_cattle_collar_2019}, and the legs using cuffs \cite{tran_animal_behaviour_recognition_2022}. 

In general, there are two main areas of operation for cattle trackers.
Indoor tracking systems mainly focus on the aspect of health monitoring and optimizing the farming process, such as automated milking, feeding, or calving control \cite{calving_indoor_tracking_2020, LELY_solutions, SMAXTEC_leistungsumfang}. 
However, when it comes to outdoor trackers, the application scenario is different. 
On large pastures, the focus mainly lies on the localization task. 
As the data throughput is limited by long-range compatible protocols, such as \ac{LoRa}, and a long battery lifetime is required, the transmission frequency is kept as low as possible and additional information transfer in the payload is usually restricted to light-weight event reporting \cite{facina_cattle_theft_2022, jegan_cattle_tracking_2022}.

Singhal \textit{et al.} \cite{singhal_cattle_collar_2022} proposes a cattle collar, equipped with an accelerometer, magnetometer, and gyroscope to classify the cow's activity to the three labels rumination, eating, and no jaw movement.
The collected data is sent over Wi-Fi to a back-end server where a random forest classifier predicts the current activity.

In \cite{tran_animal_behaviour_recognition_2022}, the authors proposed a combination of two trackers with an accelerometer, one mounted on the leg and one on the collar. The collected data are sent over \ac{LoRa} to a back-end server where the cow's actual behavior is classified as either feeding, lying, standing, or walking using a random forest algorithm. 
Although both examples achieve a good performance in their predictions, they rely on constant data transmission as the data evaluation is only made on the back-end server. This affects the overall battery lifetime of the sensor nodes.
Having an intelligent algorithm, or even a small \ac{ML} algorithm helps to reduce the amount of data to be transmitted and therefore can contribute to improving the energy efficiency by reducing the transmitted data volume \cite{magno_wulora_2017}.

In addition to research-based devices, there also exist various industrial variants that are battery-powered or integrate solar energy harvesting.
\cite{MOOVEMENT_gpseartag, HALTER_smartcollar, DIGITANIMAL_gpscattletracker, DIGITALMATTER_Yabby3}.
The commercial product in \cite{LONESTAR_gpscattletracking} has a similar design and shape as our proposed system. Running from three non-rechargeable batteries without energy harvesting, it achieves an estimated lifetime of 11 weeks with an update rate of 15 minutes. Integrating energy harvesting, either from kinetic or solar sources, is essential as would extend the rather limited lifetime and reduce maintenance \cite{magno2018micro}.

\begin{figure}[t]
    \centering
    \includegraphics[width=\columnwidth]{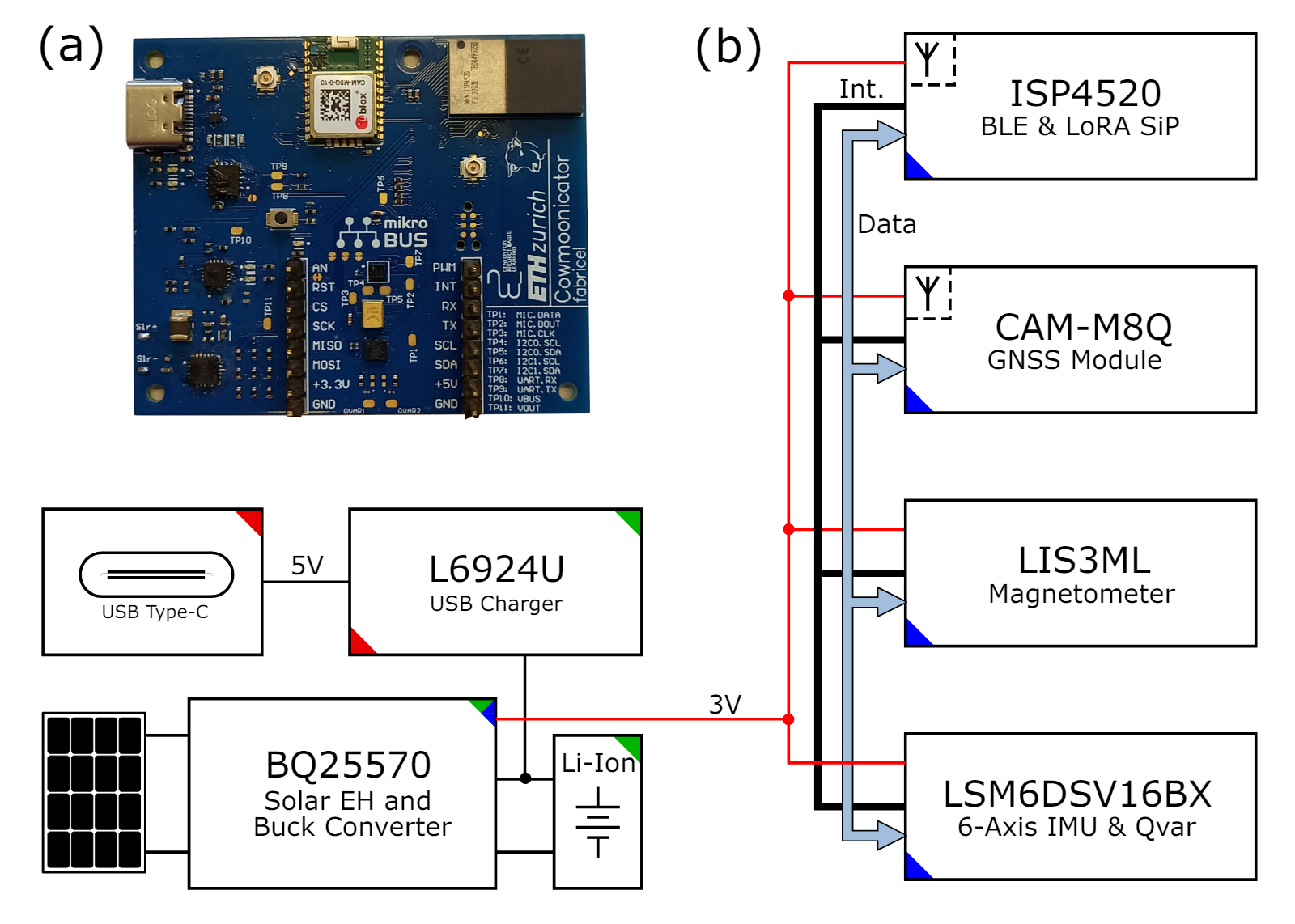}
    \vspace{-7mm}
    \caption{(a): Illustration of the custom-designed sensor board. (b): High-level architecture of the proposed sensor node, divided into the three main sections; communication and processing, activity tracking, and power subsystem.}
    \label{fig:hardware_architecture}
\end{figure}
\section{System Definition}\label{sec:system_adefinition}
The proposed system for cattle tracking consists of both, hardware and software components.
Trackers around the cows' necks serve as edge devices, gathering, preprocessing, and transmitting the measurements via \ac{LoRaWAN} to a specific gateway. 
From there, the received packets are forwarded to a central server, which decodes the payload and stores the collected data in a MySQL database, for further processing (\autoref{fig:system_overview}).

\subsection{Hardware Architecture}\label{subsec:hardware_architecture}

The cattle tracker has been designed to continuously collect data about the cow's activity and position while being on an alpine pastures. As these pastures are often in remote areas with less infra-structure, data is transmitted via \ac{LoRaWAN}, requiring the sensor data to be pre-processed onboard and with low energy expenditure. Therefore the following sensors have been integrated to determine the cow's activity parameters:
\begin{itemize}
    \item \textit{Accelerometer:} It allows recognizing the cattle's activities based on movement events, such as walking, lying down, and standing up. 
    From this, a generalized statement on the cow's activity can be derived.
    \item \textit{Magnetometer:} Since the cattle tracker has been placed around the cow's neck, a magnetometer can give information about the relative head tilt. Based on the determined angle, the grazing time can be determined. 
    \item \textit{GNSS module:} Having the GNSS position of each individual cow in a herd facilitates livestock control and simplifies the cattle collection on alpine pastures. In combination with the other sensors, more elaborated information, such as herd- and social behavior, preferred grazing spots, and traveled distance can be derived.
\end{itemize}

The designed cattle tracker can be divided into three main parts: 
communication and processing, activity tracking, and power subsystem, \autoref{fig:hardware_architecture}. 
Commercial off-the-shelf components reduce the overall cost per node while giving flexibility for custom adaptations and extensions.

\textbf{\textit{Communication and Processing:}}
The cattle tracker is built around the \ac{SiP} \textit{ISP4520-EU} (Insight SIP). 
The module combines a low power \ac{BLE} enabled microcontroller, namely an nRF52832 \ac{SoC} (Nordic Semiconductor), with down to \qty{3.4}{\micro\watt} waiting on RTC as well as the sub-GHz radio transceiver \textit{SX1261} (Semtech) in a single package.
Further, it integrates \ac{RF} matching, as well as an optional internal antenna for both wireless protocols and both, a \qty{32}{\kilo\hertz} and \qty{32}{\mega\hertz} crystal, reducing the needs of external components to a minimum and reducing sensor node cost.
The \ac{GNSS} position is acquired using a \textit{CAM-M8Q} concurrent standard precision \ac{GNSS} module (ublox) with an integrated chip antenna, allowing for a fully closed housing with no protruding antennas. However, to increase the link budget for the LoRa communication, and to increase the GNSS sensitivity, larger antennas with a higher gain (such as patch antennas) can be connected over an optional standardized \qty{50}{\ohm} U.FL connector and used instead of the module's internal antenna.

\textbf{\textit{Activity Tracking:}}
An \ac{IMU} of type \textit{LSM6DSV16BX} (ST Microelectronics) has been integrated to detect the cow's movements. It was chosen due to low energy consumption during operation (\qty{36}{\micro\watt} at \qty{60}{\hertz}), enabling continuous monitoring for movement events. Similarly, to determine the inclination of the head, the magnetometer \textit{LIS3ML} (ST Microelectronics) has been selected
Both sensors feature dedicated interrupt lines to alert the \ac{MCU} in case of a pre-defined event.

\textbf{\textit{Power Subsystem:}}
The whole system is powered by a rechargeable Li-ion battery with a total capacity of \qty{5.2}{\Ah} and can be charged in two ways: it can be charged over a USB type-C connector and battery charger
system \textit{L6924U} (ST Microelectronics), or over the ultra-low power energy harvester and power management \ac{IC} \textit{BQ25570} (Texas Instruments).
The latter also generates the system's voltage of \qty{3}{\volt} directly from the battery.
Four separated solar cells of type \textit{KXOB25-14X1F} (Anysolar) with a total area of \qty{7.36}{\centi\meter\squared} have been integrated to harvest energy from solar radiation. The separation into 4 cells improves mechanical resilience in case of collisions while keeping the antenna areas free for optimum \ac{GNSS} and \ac{LoRaWAN} reception. The solar cell size is calculated as a trade-off between battery size and solar cell size,  keeping the dimension of the tracker as small as possible, while allowing sufficient power to operate over the alpine period and maintain a physical robustness for deployment (see also Section \ref{subsec:results_energy}).

\subsection{Hardware Integration} \label{subsec:hardware_integration}
\begin{figure}[t]
    \centering
    \includegraphics[width=\columnwidth]{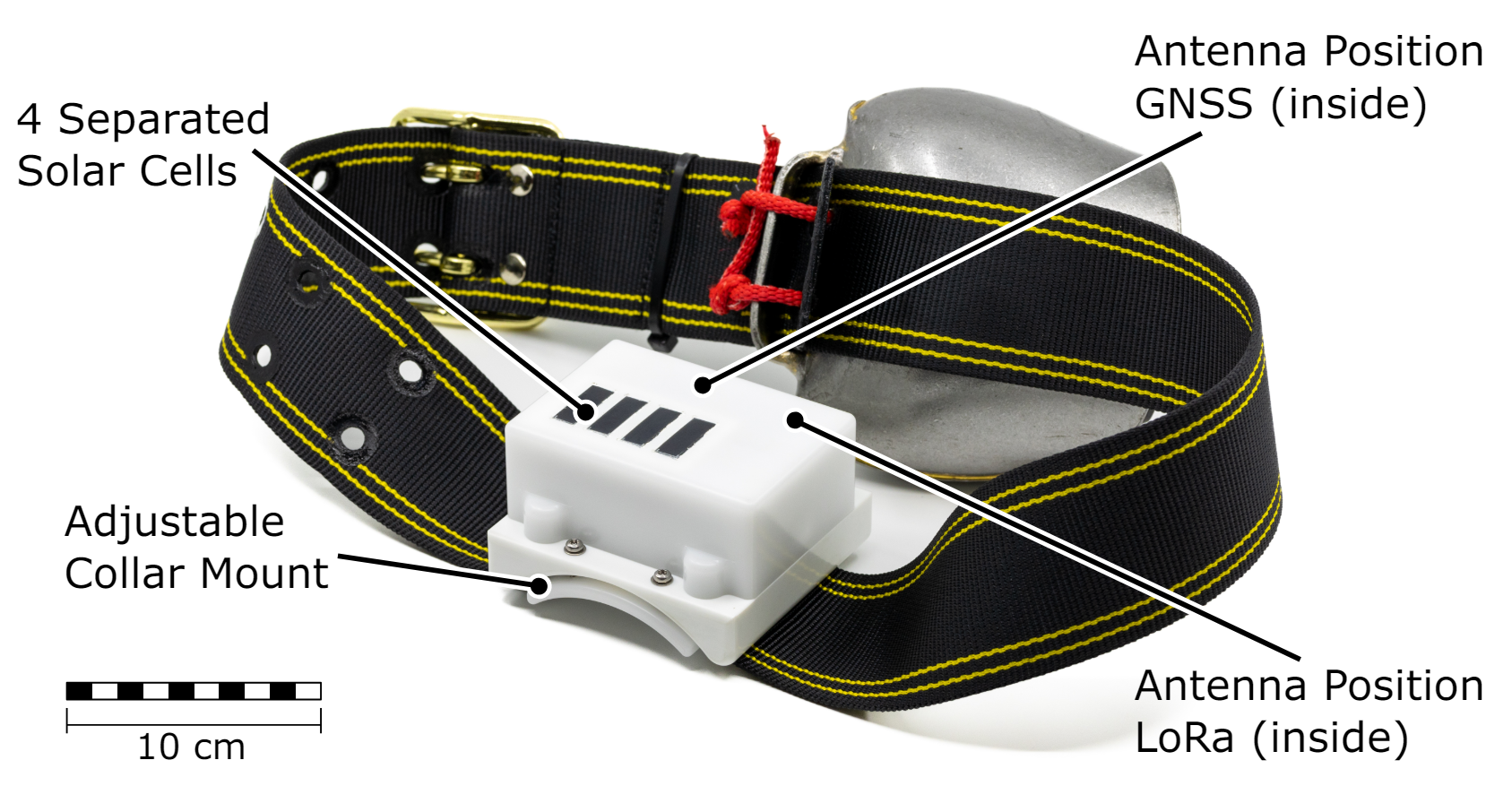}
    \vspace{-5mm}
    \caption{Fully-assembled cow tracker, mounted to a collar with a cowbell. The dimensions of the trackers are 90 mm by 85 mm by 54 mm and achieved a total weight of \qty{1.29}{\kilo\gram}.}
    \label{fig:assembled_hardware}
\end{figure}

The hardware of the proposed cow tracker is designed to accommodate multiple sensors, microcontroller, and battery in a custom-designed robust, and weatherproof housing made of \ac{POM}, a high-quality thermoplastic material that is well suited to withstand harsh environments due to its high rigidity and stability.
Rounded corners make sure there is no chance of injury while wearing. The device supports collars with a maximal width of \qty{65}{\milli\meter} by clamping them between the tracker itself and its curved lower part, \autoref{fig:assembled_hardware}.
With the dimensions of \qty{90}{\milli\meter} in length, \qty{85}{\milli\meter} in width, and \qty{54}{\milli\meter} in height, the tracker achieves a weight of \qty{382}{\gram}. Together with the bell (Firmann) and collar, the whole system reaches a total weight of \qty{1.29}{\kilo\gram}. Compared to large traditional cow bells that can achieve weights of over \qty{5.5}{\kilo\gram} \cite{johns_cow_2017}, the proposed cow tracker is much lighter and does thus not affect the cow's natural behavior. 
As the tracker and the bell solve the same task of simplifying cow localization, the bell could be omitted.

\subsection{Firmware Architecture}\label{subsec:firmware_architecture}

\begin{figure}[t]
    \centering
    \includegraphics[width=\columnwidth]{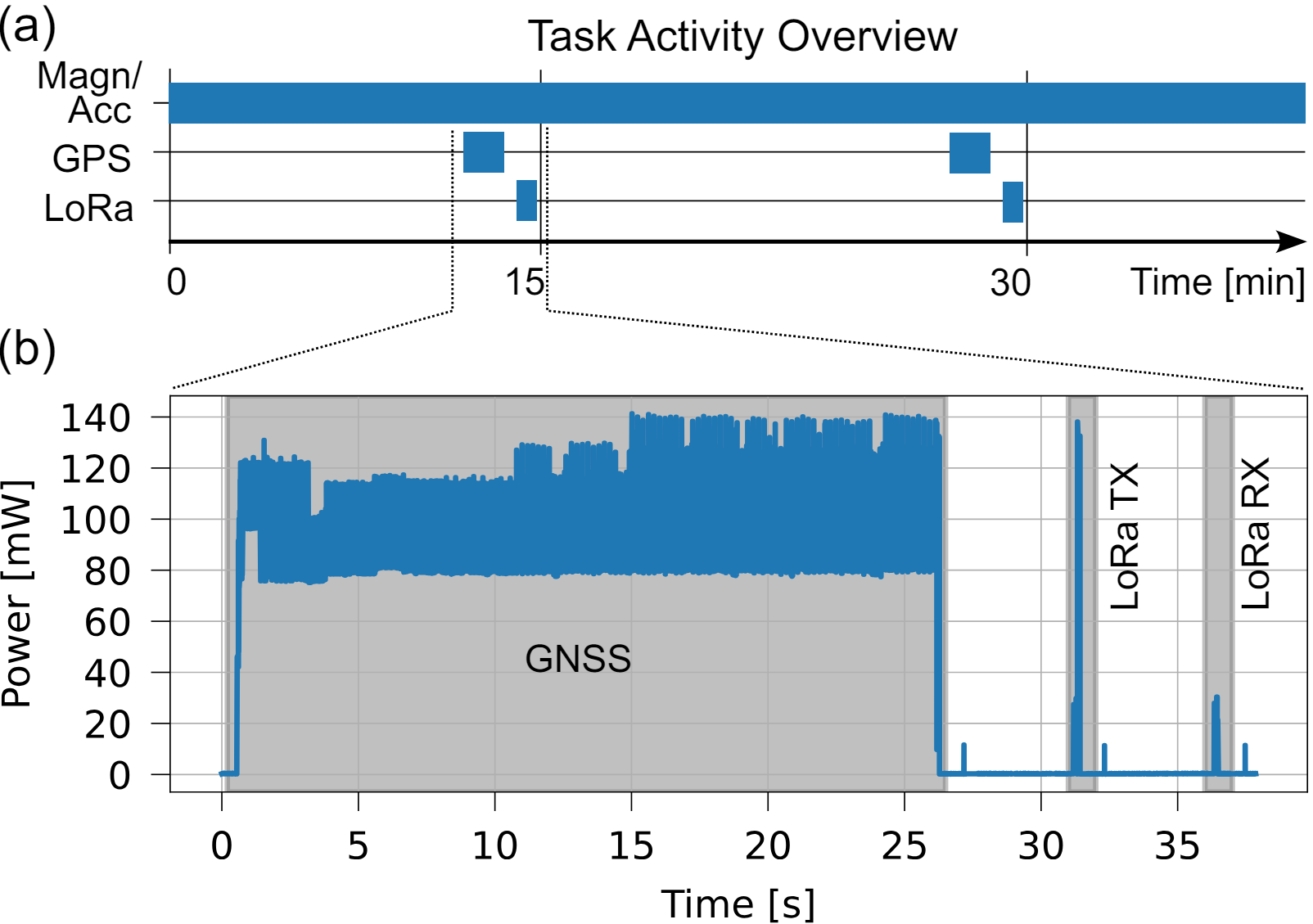}
    \vspace{-7mm}
    \caption{Concept drawing of the individual task activation as orchestrated by firmware (a) and detailed power draw on a battery of worst-case event with \ac{GNSS} not locked (b).}
    \label{fig:powerprofile}
\end{figure}

The firmware is based on Zephyr \ac{RTOS}, organized into different software tasks: (1) the main task handles the system setup and sensor configuration, key exchanges with the \ac{LoRaWAN} gateway as well as the data aggregation into a \ac{LoRa} packet and transmission of the data, (2) the \ac{GNSS} task regularly polls the \ac{GNSS} module and parses the output for further transmission and (3) the magnetometer/accelerometer task continuously samples magnetometer and accelerometer and calculates the orientation of the device with respect to ground and magnetic north. The periodicity of the tasks can be freely chosen in the firmware, for the results presented in section \ref{subsec:results_activity} it was set to \qty{5}{\minute}, for general operation on the pasture a baseline of \qty{15}{\minute} is considered as a trade-off between power draw and measurement resolution for both, the \ac{GNSS} and LoRa operation. A coarse overview of the task scheduling is given in \autoref{fig:powerprofile} (a).

In order to save transmission energy, the data acquired by the magnetometer, \ac{IMU} and \ac{GNSS} is pre-processed onboard before transmitting. The device orientation is calculated based on the accelerometer and gyroscope readings, low-pass filtered by an \ac{IIR}-filter with a corner frequency of approx. \qty{0.1}{\hertz}. The orientation with respect to magnetic north is calculated by removing the magnetometer offset and calculating the arc-tangent of the magnetic field vector.

The onboard pre-processed data is packed into frames of 31 bytes in total. Among debug and timestamp information, the \ac{GNSS} position is transmitted as 4-byte values, including timestamp and estimated accuracy as given by the \ac{GNSS} module. The device orientation is transmitted as 1 1-byte rotation value (pointing North) and 2-byte movement average detected by the \ac{IMU}. The air-time of these packets is dependent on the LoRaWAN settings, and the payload \cite{semtech_2013}. For the given spreading factor of 8 (trade-off between range and airtime), the 31 byte requires approximately \qty{170}{\milli\second} of airtime.

\begin{table}[b]
    \centering
    \caption{Energy consumption of sensors per hour}
    \label{tab:energy}
    \begin{tabular}{p{2.5cm} p{2.5cm} p{2.5cm}}
        \hline
        {\bf Sensor}        &   {\bf active time}           &   {\bf Energy}                          \\      \hline
        LoRaWAN (SF7)       & \qty{0.2}{\second\per\hour}   &   \SI{24}{\milli\joule\per\hour}       \\      
        GNSS                & 4x \qty{25}{\second\per\hour} &   \SI{11.53}{\joule\per\hour}        \\      
        Accelerometer       & cont. (\qty{60}{\hertz})      &   \SI{216}{\milli\joule\per\hour}       \\      
        Magnetometer        & cont. (\qty{20}{\hertz})      &   \SI{2.43}{\joule\per\hour}     \\      
        MCU                 & \qty{70}{\second\per\hour}    &   \SI{7.13}{\joule\per\hour} \\\hline
        \textbf{Total}      &                               &   \textbf{\SI{511.9}{\joule} per day}   \\   \hline   
    \end{tabular}
\end{table}

\begin{figure}[hb!]
   \centering
    \includegraphics[width=\columnwidth]{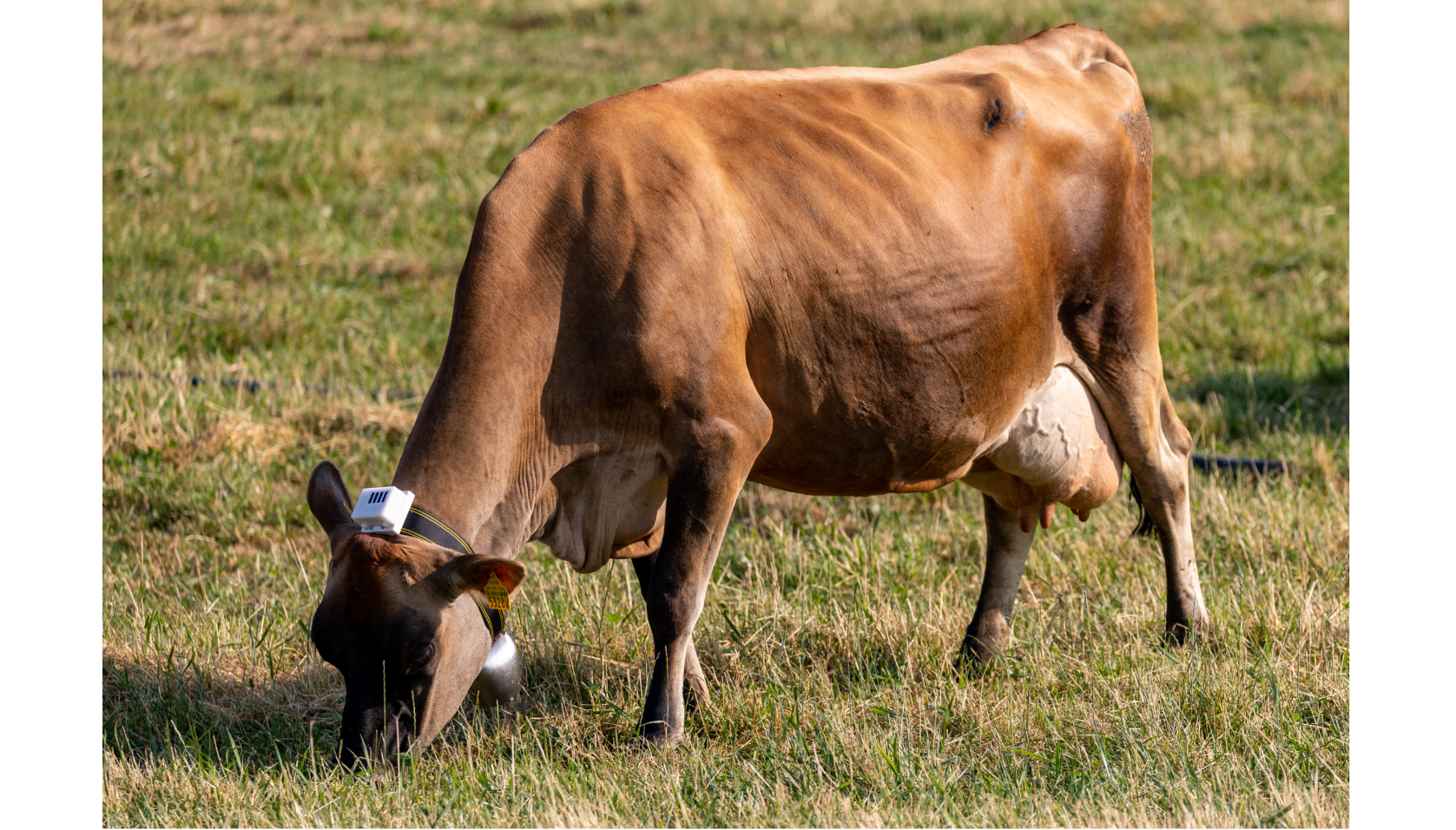}
    \vspace{-5mm}
    \caption{The proposed cow tracker is attached to the bell collar and worn by a cow during the evaluation phase.
    The position and geometry of the tracker are designed to reduce the movement during the deployment, allowing to collect data with minimal random movement artifacts.
    }
    \label{fig:tracker_on_cow}
\end{figure}

\section{System Evaluation}\label{sec:system_evaluation}

\begin{figure*}[ht]
    \centering
    \includegraphics[width=\textwidth]{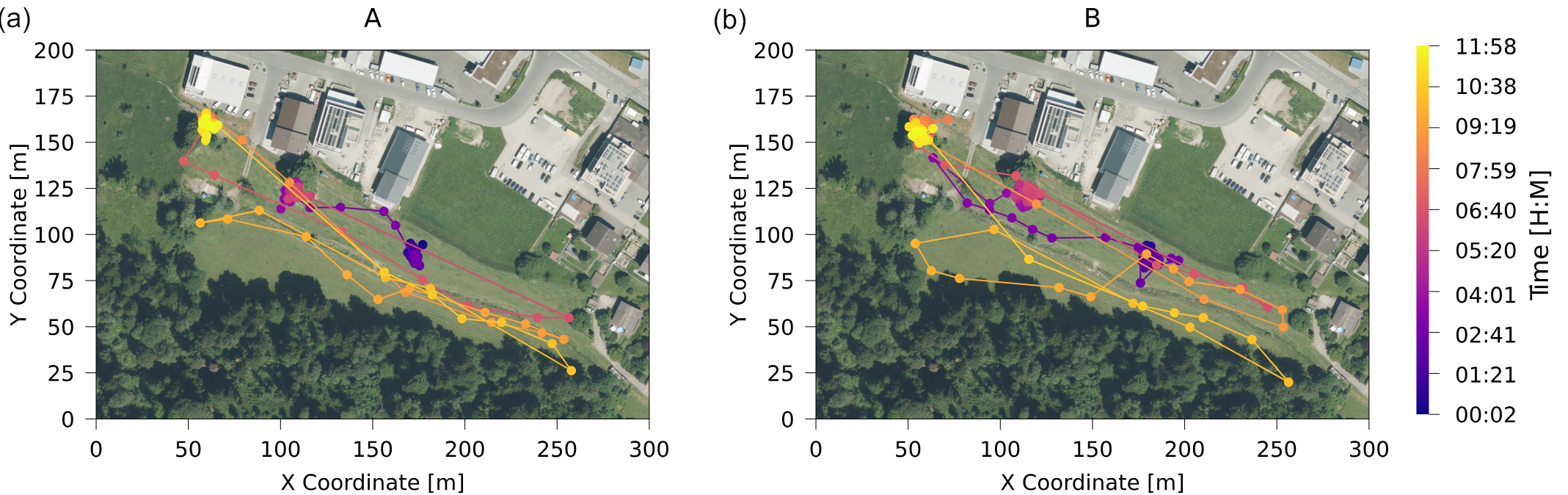}
    \vspace{-7mm}
    \caption{Comparison of the individual's position over 12 hours on a pasture with a length of \qty{275}{\meter} and a total area of approximately \qty{20280}{\meter\squared}. 
   A, as well as B, show similar walking patterns. They rest on similar positions at the same time. Map source: \cite{orthofoto}
    }
    \label{fig:cowtrackingresults}
\end{figure*}

The evaluation of the overall system is based on two main parts: 
\begin{enumerate}
    \item Evaluating the tracing systems performance, energy consumption, and achieved energy harvesting. Further, a theoretical assessment of the expected lifetime is made.
    \item Assessing the process of data collection, extrapolate activity parameters using sensor fusion to lay the foundation for possible on-board \ac{ML} algorithms for future work.
\end{enumerate}


\subsection{Energy Harvesting and Self-Sustainability}\label{subsec:results_energy}
In order to design a fully self-sustained system over the alpine season which lasts 5 months, starting end of April and ending in October, the power consumption needs to be well-balanced with the available energy resources. To relax the power budget, and to increase the battery lifetime, the proposed device integrates four solar cells of type \textit{KXOB25-14X1F} (Anysolar) with a total area of \qty{7.36}{\centi\meter\squared}.
The solar energy calculations are based on a long-term energy harvesting evaluation with using the implemented energy harvesting sub-circuitry.
This experimental evaluation of continuous data acquisition for over 1 month has been conducted to evaluate the energy harvesting subsystem.
The solar cells have been mounted at an angle of \qty{30}{\degree} with respect to the horizontal plane to take the tracker's shift during the deployment into account.

Over the whole period of time, the solar cells were able to harvest \qty{184.9}{\joule} per day. 
Compared to the harvesting, the energy required by the device is presented in \autoref{tab:energy}. 
Among the largest contributors is the always-on magnetometer with a sampling rate of \qty{20}{\hertz} (\SI{2.43}{\joule\per\hour}), as well as the \ac{MCU} in order to process the data stream of these sensors with approx. \qty{7.13}{\joule\per\hour} and the \ac{GNSS} module which determines the tracker's position every 15 minutes.
Based on all individual contributions, the system consumes approx. \qty{511.9}{\joule} per operational day. 
This, together with the energy available from an integrated battery with a capacity of \qty{5.2}{\ampere\hour} is enough to be maintenance-free for four months.
However, by increasing the area covered by solar cells from \qty{7.36}{\centi\meter\squared} by a factor of 2.78 to \qty{20.46}{\centi\meter\squared}, enough energy can be harvested to reach full self-sustainability. With an upper side surface area of \qty{48.6}{\centi\meter\squared}, the proposed tracker has enough space to accommodate the enlarged solar cells.


\subsection{Activity Tracking}\label{subsec:results_activity}
The feasibility of the designed sensor system is presented by tracking cattle activity within an open pasture over a period of 12 hours to 3 days. For this, two of the sensing devices (A and B, \autoref{fig:tracker_on_cow}) were mounted on cattle in a herd. Data from the sensing devices was onboard processed and the results were transmitted via the \ac{LoRaWAN} connection to a gateway in the vicinity of the pasture and stored in the server database for later evaluation. Alternatively, a dashboard could show the current status and measurements of each device.
As the first indicator of cattle activity, their position within the pasture over time is evaluated. The employed \ac{GNSS} module uBlox \textit{CAM-M8Q} module offers a precision of approx. \qty{2.5}{\meter} \cite{noauthor_datasheet_2023}, which is similar to the size of the cattle, and thus sufficient for coarse localization in an open pasture. 
The individual \ac{GNSS} position estimates are shown in \autoref{fig:cowtrackingresults}, with the timestamp of each position coded in the color map. The example measurement starts at midnight and is then displayed for 12 consecutive hours. 
Noticeably, both cattle rest over longer periods of time in the same position and then move to a new one. In the presented data set, these main resting places are utilized between midnight and approximately 4 a.m. From early morning until 12 p.m., significantly more movement across the accessible area is visible. 
Next to the overall time spent in different locations of the pasture, not all areas were visited by the cattle, especially in the southwest of the fenced area. Comparing the measurements from the two devices, the resting places, as well as movement, are similar, indicating a matching walking pattern between the individual cattle. 

\begin{figure}[hb!]
    \centering
    \includegraphics[width=\columnwidth]{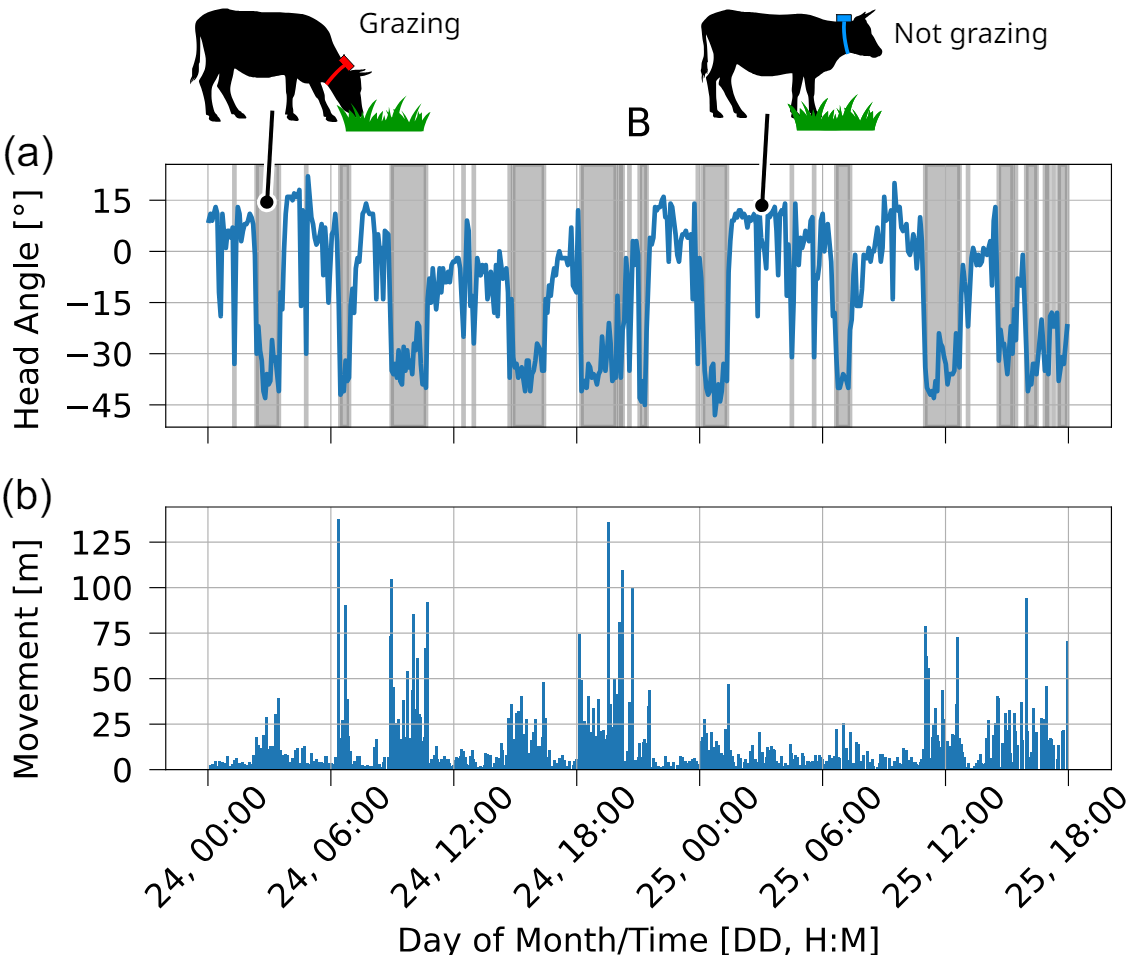}
    \vspace{-5mm}
        \caption{Activity comparison with walked distance over two days, showing the individual's Relative angle obtained by the magnetometer's x-axis and the traveled distance obtained from GPS coordinates. Angles below the threshold of \qty{-20}{\degree} indicate grazing, whereas angles above \qty{-10}{\degree} indicate not grazing. Grazing is marked in gray on the plot.}
    \label{fig:head_distance_b}
\end{figure}

Within the same data recording, evaluating the accelerometer and magnetometer data as a second indicator of activity is possible. This yields the relative angle of the cattle's head with respect to earth, allowing simple behavior activity detection. As validated by manual observation during \qty{4}{\hour}, this value indicates if the cattle has its head lifted, is resting, or grazing with the head down on the grass \autoref{fig:head_distance_b} (a). 
By manual observation of the cattle wearing the sensor devices, a head angle threshold below \qty{-20}{\degree} was defined such that the cow can be classified as grazing, with angles above \qty{-10}{\degree} classified as not grazing. These regions are marked within \autoref{fig:head_distance_b} as gray-shaded areas. Possible applications for these estimates would be the overall grazing time per cattle, calculated to be \qty{12.6}{\hour}  for cattle A, and \qty{13.8}{\hour} for cattle B during the course of the experiment. Noticeably, also during night, the devices indicate the cattle grazing, in between periods of non-grazing. 

\begin{figure}[ht!]
    \centering
    \includegraphics[width=\columnwidth]{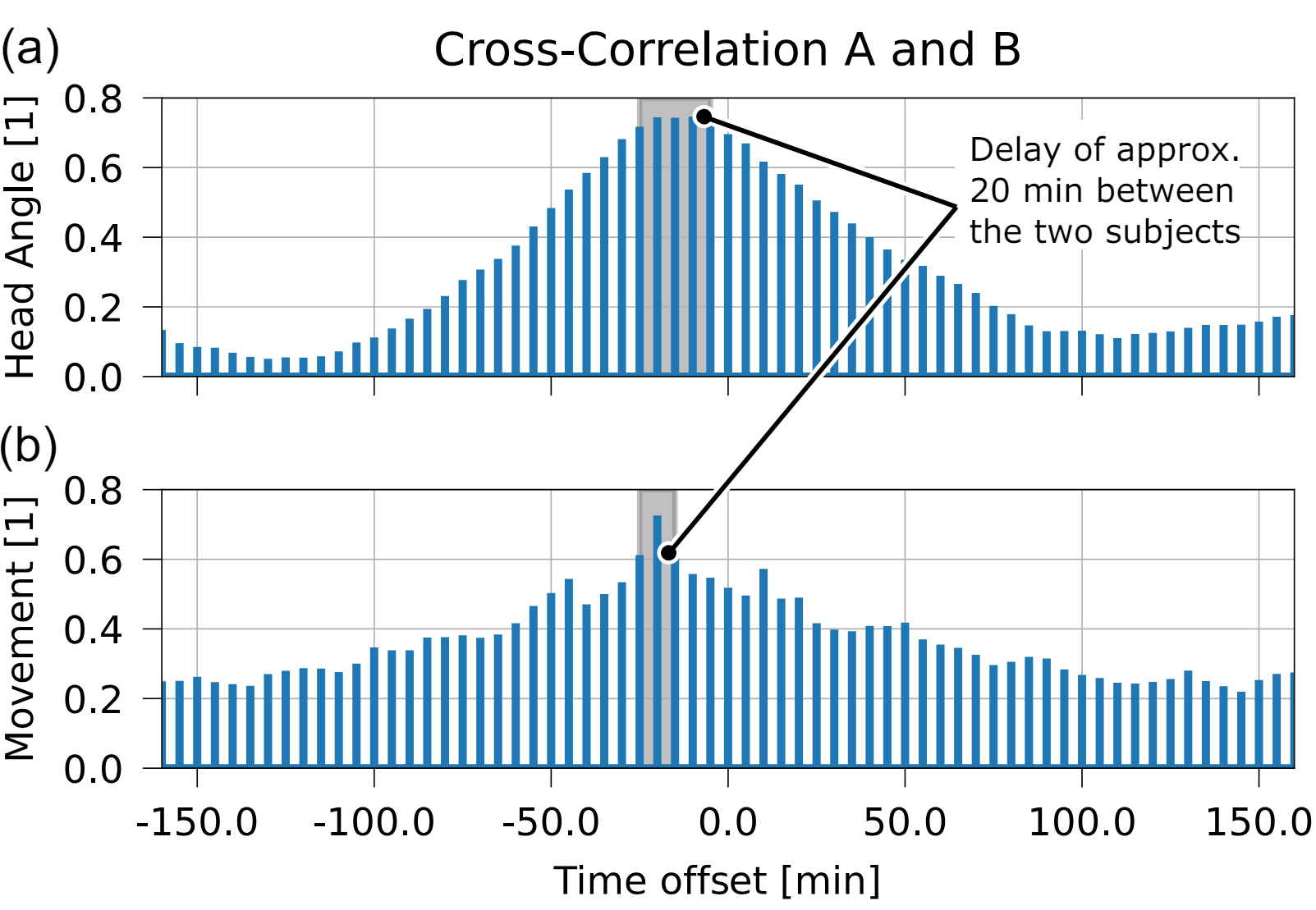}
    \vspace{-8mm}
    \caption{Cross correlation between the head angle (a) and movement (b) of subjects A and B, with a correlation factor of up to 0.7 at a lag of \qty{20}{\minute} between each other.}
    \label{fig:crosscorr_angle_distance}
\end{figure}

Based on the concurrent \ac{GNSS} position recording with respect to the head angle calculations, the average movement per time-interval (\qty{5}{\minute}) across the pasture is displayed in \autoref{fig:head_distance_b} (b). When grazing, the traveled distance is significantly higher, with up to \qty{130}{\meter} traveled within \qty{5}{\minute} and an average movement of approx. \qty{25}{\meter}. When not grazing, there is also movement detected, which is in part due to noise of the \ac{GNSS} measurement (specified in \cite{noauthor_datasheet_2023}) as well as small movements of the cattle. Similar to the results of \autoref{fig:cowtrackingresults}, the traveled distance, as well as the accumulated distances, correlate well over time between the two devices, indicating similar movements of the individual cattle during this measurement.

Evaluating the movement and head angle recorded by devices A and B, a noticeable degree of similarity is observed. 
The similarity is further investigated by cross-correlation of the movement across the devices, as well as the head angle data. This yields a correlation factor of up to 0.68 for the movement, with a delay below \qty{20}{\minute} between devices, as seen in \autoref{fig:crosscorr_angle_distance} (b). 
This implies the two cattle are following each other in terms of movement. 
Similarly, the head angle correlation factor between the devices reaches up to 0.75, as presented in \autoref{fig:crosscorr_angle_distance} (b) with the same lag of approx. \qty{20}{\minute}.

\newpage
\section{Conclusion}\label{sec:conclusion}
This paper presented the design, implementation, and in-field evaluation of maintenance-free, long-range wireless intelligent sensors for cow tracking on alpine pastures, collecting activity parameters to assess their health status.
Leveraging energy harvesting from small-size solar panels, the achieved lifetime from a full battery without recharging with data acquisition and transmission over \ac{LoRa} on a 15-minute basis is over 4 months, eliminating maintenance and allowing simple deployment.
In a real-field use case, we demonstrate that the energy harvesting subsystem increased the lifetime by 40\% to 6 months.
When increasing the solar cells' area by a factor of 2.78 from \qty{7.36}{\centi\meter\squared} to \qty{20.46}{\centi\meter\squared}, we estimate that the system can achieve self-sustainability. 
In-field experimental evaluation of three days of data from two cows indicates the potential knowledge gained by employing such livestock trackers. the activity \textit{grazing} can be clearly distinguished by applying a simple static threshold. 

\begin{acronym}
    \acro{RF}{Radio Frequency}
    \acro{IoT}{Internet of Things}
    \acro{IoUT}{Internet of Underwater Things}
    \acro{UWN}{Underwater Wireless Network}
    \acro{UWSN}{Underwater Wireless Sensor Node}
    \acro{AUV}{Autonomous Underwater Vehicles}
    \acro{UAC}{Underwater Acoustic Channel}
    \acro{FSK}{Frequency Shift Keying}
    \acro{OOK}{On-Off Keying}
    \acro{ASK}{Amplitude Shift Keying}
    \acro{UUID}{Universal Unique Identifier}
    \acro{PZT}{Lead Zirconium Titanate}
    \acro{AC}{Alternating Current}
    \acro{NVC}{Negative Voltage Converter}
    \acro{NVCR}{Negative Voltage Converter Rectifier}
    \acro{FWR}{Full-Wave Rectifier}
    \acro{MCU}{Microcontroller}
    \acro{GPIO}{General Purpose Input/Output}
    \acro{PCB}{Printed Circuit Board}
    \acro{AUV}{Autonomous Underwater Vehicle}
    \acro{IMU}{Intertial Measurement Unit}
    \acro{BLE}{Bluetooth Low Energy}
    \acro{FSR}{Force Sensing Resistor}
    \acro{SiP}{System in Package}
    \acro{SoC}{System on Chip}
    \acro{SpO2}{Oxigen Saturation}
    \acro{PULP}{Parallel Ultra-Low Power}
    \acro{ML}{Machine Learning}
    \acro{ADC}{Analog to Digital Converter}
    \acro{TCDM}{Tightly Coupled Data Memory}
    \acro{GNSS}{Global Navigation Satellite System}
    \acro{IC}{Integrated Circuit}
    \acro{POM}{Polyoxymethylene}
    \acro{RTOS}{Real-Time Operating System}
    \acro{LoRa}{Long Range}
    \acro{LoRaWAN}{Long Range Wide Area Network}
    \acro{ML}{Machine Learning}
    \acro{IIR}{Infinite Impulse Response}
\end{acronym}

\newpage 
\bibliographystyle{ACM-Reference-Format}
\bibliography{cow_tracking}


\begin{thebibliography}{43}


\ifx \showCODEN    \undefined \def \showCODEN     #1{\unskip}     \fi
\ifx \showDOI      \undefined \def \showDOI       #1{#1}\fi
\ifx \showISBNx    \undefined \def \showISBNx     #1{\unskip}     \fi
\ifx \showISBNxiii \undefined \def \showISBNxiii  #1{\unskip}     \fi
\ifx \showISSN     \undefined \def \showISSN      #1{\unskip}     \fi
\ifx \showLCCN     \undefined \def \showLCCN      #1{\unskip}     \fi
\ifx \shownote     \undefined \def \shownote      #1{#1}          \fi
\ifx \showarticletitle \undefined \def \showarticletitle #1{#1}   \fi
\ifx \showURL      \undefined \def \showURL       {\relax}        \fi
\providecommand\bibfield[2]{#2}
\providecommand\bibinfo[2]{#2}
\providecommand\natexlab[1]{#1}
\providecommand\showeprint[2][]{arXiv:#2}

\bibitem[AlZubi and Galyna(2023)]%
        {alzubi_smart_farming_2023}
\bibfield{author}{\bibinfo{person}{Ahmad~Ali AlZubi} {and} \bibinfo{person}{Kalda Galyna}.} \bibinfo{year}{2023}\natexlab{}.
\newblock \showarticletitle{Artificial Intelligence and Internet of Things for Sustainable Farming and Smart Agriculture}.
\newblock \bibinfo{journal}{\emph{IEEE Access}}  \bibinfo{volume}{11} (\bibinfo{year}{2023}), \bibinfo{pages}{78686--78692}.
\newblock
\urldef\tempurl%
\url{https://doi.org/10.1109/ACCESS.2023.3298215}
\showDOI{\tempurl}


\bibitem[Benaissa et~al\mbox{.}(2020)]%
        {calving_indoor_tracking_2020}
\bibfield{author}{\bibinfo{person}{S. Benaissa}, \bibinfo{person}{F.A.M. Tuyttens}, \bibinfo{person}{D. Plets}, \bibinfo{person}{J. Trogh}, \bibinfo{person}{L. Martens}, \bibinfo{person}{L. Vandaele}, \bibinfo{person}{W. Joseph}, {and} \bibinfo{person}{B. Sonck}.} \bibinfo{year}{2020}\natexlab{}.
\newblock \showarticletitle{Calving and estrus detection in dairy cattle using a combination of indoor localization and accelerometer sensors}.
\newblock \bibinfo{journal}{\emph{Computers and Electronics in Agriculture}}  \bibinfo{volume}{168} (\bibinfo{year}{2020}), \bibinfo{pages}{105153}.
\newblock
\showISSN{0168-1699}
\urldef\tempurl%
\url{https://doi.org/10.1016/j.compag.2019.105153}
\showDOI{\tempurl}


\bibitem[Cabezas et~al\mbox{.}(2022)]%
        {cabezas_cattle_behaviour_2022}
\bibfield{author}{\bibinfo{person}{Javier Cabezas}, \bibinfo{person}{Roberto Yubero}, \bibinfo{person}{Beatriz Visitación}, \bibinfo{person}{Jorge Navarro-García}, \bibinfo{person}{María~Jesús Algar}, \bibinfo{person}{Emilio~L. Cano}, {and} \bibinfo{person}{Felipe Ortega}.} \bibinfo{year}{2022}\natexlab{}.
\newblock \showarticletitle{Analysis of Accelerometer and GPS Data for Cattle Behaviour Identification and Anomalous Events Detection}.
\newblock \bibinfo{journal}{\emph{Entropy}} \bibinfo{volume}{24}, \bibinfo{number}{3} (\bibinfo{year}{2022}).
\newblock
\urldef\tempurl%
\url{https://doi.org/10.3390/e24030336}
\showDOI{\tempurl}


\bibitem[Chapa et~al\mbox{.}(2020)]%
        {jose_animal_welfare_2020}
\bibfield{author}{\bibinfo{person}{Jose~M. Chapa}, \bibinfo{person}{Kristina Maschat}, \bibinfo{person}{Michael Iwersen}, \bibinfo{person}{Johannes Baumgartner}, {and} \bibinfo{person}{Marc Drillich}.} \bibinfo{year}{2020}\natexlab{}.
\newblock \showarticletitle{Accelerometer systems as tools for health and welfare assessment in cattle and pigs – A review}.
\newblock \bibinfo{journal}{\emph{Behavioural Processes}}  \bibinfo{volume}{181} (\bibinfo{year}{2020}), \bibinfo{pages}{104262}.
\newblock
\urldef\tempurl%
\url{https://doi.org/10.1016/j.beproc.2020.104262}
\showDOI{\tempurl}


\bibitem[Confederation(2022)]%
        {alpine_season_2022}
\bibfield{author}{\bibinfo{person}{Swiss Confederation}.} \bibinfo{year}{2022}\natexlab{}.
\newblock \bibinfo{title}{Living traditions - the Alpine pasture season}.
\newblock
\newblock


\bibitem[Corporation(2013)]%
        {semtech_2013}
\bibfield{author}{\bibinfo{person}{Semtech Corporation}.} \bibinfo{year}{2013}\natexlab{}.
\newblock \bibinfo{title}{{AN1200}.13 {SX1272}/3/6/7/8: {LoRa} {Modem} {Designer}’s {Guide} {Revision} 1}.
\newblock
\newblock


\bibitem[Di~Nuzzo et~al\mbox{.}(2021)]%
        {nuzzo_structural_health_2021}
\bibfield{author}{\bibinfo{person}{Flavio Di~Nuzzo}, \bibinfo{person}{Davide Brunelli}, \bibinfo{person}{Tommaso Polonelli}, {and} \bibinfo{person}{Luca Benini}.} \bibinfo{year}{2021}\natexlab{}.
\newblock \showarticletitle{Structural Health Monitoring System With Narrowband IoT and MEMS Sensors}.
\newblock \bibinfo{journal}{\emph{IEEE Sensors Journal}} \bibinfo{volume}{21}, \bibinfo{number}{14} (\bibinfo{year}{2021}), \bibinfo{pages}{16371--16380}.
\newblock
\urldef\tempurl%
\url{https://doi.org/10.1109/JSEN.2021.3075093}
\showDOI{\tempurl}


\bibitem[digital matter(2023)]%
        {DIGITALMATTER_Yabby3}
\bibfield{author}{\bibinfo{person}{digital matter}.} \bibinfo{year}{2023}\natexlab{}.
\newblock \bibinfo{title}{Yabby3 for LoRaWAN}.
\newblock
\newblock
\urldef\tempurl%
\url{https://www.digitalmatter.com/devices/yabby3-for-lorawan/}
\showURL{%
\tempurl}
\newblock
\shownote{[Online; accessed 2-June-2023]}.


\bibitem[Digitanimal(2023)]%
        {DIGITANIMAL_gpscattletracker}
\bibfield{author}{\bibinfo{person}{Digitanimal}.} \bibinfo{year}{2023}\natexlab{}.
\newblock \bibinfo{title}{Digitanimal GPS cattle tracker}.
\newblock
\newblock
\urldef\tempurl%
\url{https://digitanimal.co.uk/product/digitanimal-gps-cattle-tracker/}
\showURL{%
\tempurl}
\newblock
\shownote{[Online; accessed 2-June-2023]}.


\bibitem[Eeshwaroju et~al\mbox{.}(2020)]%
        {eeshwaroju_drone_delivery_2020}
\bibfield{author}{\bibinfo{person}{Sreenivas Eeshwaroju}, \bibinfo{person}{Praveena Jakkula}, {and} \bibinfo{person}{Iheb Abdellatif}.} \bibinfo{year}{2020}\natexlab{}.
\newblock \showarticletitle{An IoT based Three-Dimensional Dynamic Drone Delivery (3D4) System}. In \bibinfo{booktitle}{\emph{2020 IEEE Cloud Summit}}. \bibinfo{pages}{119--123}.
\newblock
\urldef\tempurl%
\url{https://doi.org/10.1109/IEEECloudSummit48914.2020.00024}
\showDOI{\tempurl}


\bibitem[Facina et~al\mbox{.}(2022)]%
        {facina_cattle_theft_2022}
\bibfield{author}{\bibinfo{person}{Alex~R. Facina}, \bibinfo{person}{Lenin~Patricio Jiménez~Jiménez}, \bibinfo{person}{Michelle S.~P. Facina}, \bibinfo{person}{Gustavo Fraidenraich}, {and} \bibinfo{person}{Eduardo~R. De~Lima}.} \bibinfo{year}{2022}\natexlab{}.
\newblock \showarticletitle{LoRaWAN Cattle Tracking Prototype With AI-based Coverage Prediction}. In \bibinfo{booktitle}{\emph{2022 IEEE 8th World Forum on Internet of Things (WF-IoT)}}. \bibinfo{pages}{1--6}.
\newblock
\urldef\tempurl%
\url{https://doi.org/10.1109/WF-IoT54382.2022.10152029}
\showDOI{\tempurl}


\bibitem[Farooq et~al\mbox{.}(2020)]%
        {farooq_iot_agriculture_2020}
\bibfield{author}{\bibinfo{person}{Muhammad~Shoaib Farooq}, \bibinfo{person}{Shamyla Riaz}, \bibinfo{person}{Adnan Abid}, \bibinfo{person}{Tariq Umer}, {and} \bibinfo{person}{Yousaf~Bin Zikria}.} \bibinfo{year}{2020}\natexlab{}.
\newblock \showarticletitle{Role of IoT Technology in Agriculture: A Systematic Literature Review}.
\newblock \bibinfo{journal}{\emph{Electronics}} \bibinfo{volume}{9}, \bibinfo{number}{2} (\bibinfo{year}{2020}).
\newblock
\showISSN{2079-9292}
\urldef\tempurl%
\url{https://www.mdpi.com/2079-9292/9/2/319}
\showURL{%
\tempurl}


\bibitem[Halter(2023)]%
        {HALTER_smartcollar}
\bibfield{author}{\bibinfo{person}{Halter}.} \bibinfo{year}{2023}\natexlab{}.
\newblock \bibinfo{title}{Smaxtec Leistungsumfang}.
\newblock
\newblock
\urldef\tempurl%
\url{https://www.halterhq.com/}
\showURL{%
\tempurl}
\newblock
\shownote{[Online; accessed 2-June-2023]}.


\bibitem[Hořejší et~al\mbox{.}(2020)]%
        {horejsi_smart_factory_2020}
\bibfield{author}{\bibinfo{person}{Petr Hořejší}, \bibinfo{person}{Konstantin Novikov}, {and} \bibinfo{person}{Michal Šimon}.} \bibinfo{year}{2020}\natexlab{}.
\newblock \showarticletitle{A Smart Factory in a Smart City: Virtual and Augmented Reality in a Smart Assembly Line}.
\newblock \bibinfo{journal}{\emph{IEEE Access}}  \bibinfo{volume}{8} (\bibinfo{year}{2020}), \bibinfo{pages}{94330--94340}.
\newblock
\urldef\tempurl%
\url{https://doi.org/10.1109/ACCESS.2020.2994650}
\showDOI{\tempurl}


\bibitem[Jegan et~al\mbox{.}(2022)]%
        {jegan_cattle_tracking_2022}
\bibfield{author}{\bibinfo{person}{G. Jegan}, \bibinfo{person}{I.Rexiline Sheeba}, \bibinfo{person}{P.~Kavi Priya}, \bibinfo{person}{R.M. Joany}, {and} \bibinfo{person}{T. Vino}.} \bibinfo{year}{2022}\natexlab{}.
\newblock \showarticletitle{Cattle Tracking System Architecture Using LORA}. In \bibinfo{booktitle}{\emph{2022 International Conference on Power, Energy, Control and Transmission Systems (ICPECTS)}}. \bibinfo{pages}{1--4}.
\newblock
\urldef\tempurl%
\url{https://doi.org/10.1109/ICPECTS56089.2022.10046673}
\showDOI{\tempurl}


\bibitem[Johns et~al\mbox{.}(2017)]%
        {johns_cow_2017}
\bibfield{author}{\bibinfo{person}{Julia Johns}, \bibinfo{person}{Sophie Masneuf}, \bibinfo{person}{Antonia Patt}, {and} \bibinfo{person}{Edna Hillmann}.} \bibinfo{year}{2017}\natexlab{}.
\newblock \showarticletitle{regular exposure to cowbells affects the Behavioral reactivity to a noise stimulus in Dairy cows}.
\newblock \bibinfo{journal}{\emph{Frontiers in veterinary science}}  \bibinfo{volume}{4} (\bibinfo{year}{2017}), \bibinfo{pages}{153}.
\newblock


\bibitem[Joshitha et~al\mbox{.}(2021)]%
        {joshitha_lifestock_control_2021}
\bibfield{author}{\bibinfo{person}{C Joshitha}, \bibinfo{person}{P Kanakaraja}, \bibinfo{person}{Mallela~Divya Bhavani}, \bibinfo{person}{Yerramsetti Naga~Venkata Raman}, {and} \bibinfo{person}{Tadepalli Sravani}.} \bibinfo{year}{2021}\natexlab{}.
\newblock \showarticletitle{LoRaWAN based Cattle Monitoring Smart System}. In \bibinfo{booktitle}{\emph{2021 7th International Conference on Electrical Energy Systems (ICEES)}}. \bibinfo{pages}{548--552}.
\newblock
\urldef\tempurl%
\url{https://doi.org/10.1109/ICEES51510.2021.9383749}
\showDOI{\tempurl}


\bibitem[Lely(2023)]%
        {LELY_solutions}
\bibfield{author}{\bibinfo{person}{Lely}.} \bibinfo{year}{2023}\natexlab{}.
\newblock \bibinfo{title}{Lely Solutions}.
\newblock
\newblock
\urldef\tempurl%
\url{https://www.lely.com/solutions/}
\showURL{%
\tempurl}
\newblock
\shownote{[Online; accessed 2-June-2023]}.


\bibitem[Lonestar(2023)]%
        {LONESTAR_gpscattletracking}
\bibfield{author}{\bibinfo{person}{Lonestar}.} \bibinfo{year}{2023}\natexlab{}.
\newblock \bibinfo{title}{GPS Cattle Tracking}.
\newblock
\newblock
\urldef\tempurl%
\url{https://www.lonestartracking.com/gps-cattle-tracking/}
\showURL{%
\tempurl}
\newblock
\shownote{[Online; accessed 2-June-2023]}.


\bibitem[Luzern(2020)]%
        {orthofoto}
\bibfield{author}{\bibinfo{person}{Kanton Luzern}.} \bibinfo{year}{2020}\natexlab{}.
\newblock \bibinfo{title}{Orthofoto Sommer 2020}.
\newblock
\newblock
\urldef\tempurl%
\url{https://daten.geo.lu.ch/download/of20hi08_ds_v1}
\showURL{%
\tempurl}
\newblock
\shownote{Accessed: 2023-09-15}.


\bibitem[Ma et~al\mbox{.}(2020)]%
        {ma_eh_iot_2020}
\bibfield{author}{\bibinfo{person}{Dong Ma}, \bibinfo{person}{Guohao Lan}, \bibinfo{person}{Mahbub Hassan}, \bibinfo{person}{Wen Hu}, {and} \bibinfo{person}{Sajal~K. Das}.} \bibinfo{year}{2020}\natexlab{}.
\newblock \showarticletitle{Sensing, Computing, and Communications for Energy Harvesting IoTs: A Survey}.
\newblock \bibinfo{journal}{\emph{IEEE Communications Surveys \& Tutorials}} \bibinfo{volume}{22}, \bibinfo{number}{2} (\bibinfo{year}{2020}), \bibinfo{pages}{1222--1250}.
\newblock
\urldef\tempurl%
\url{https://doi.org/10.1109/COMST.2019.2962526}
\showDOI{\tempurl}


\bibitem[Magno et~al\mbox{.}(2017)]%
        {magno_wulora_2017}
\bibfield{author}{\bibinfo{person}{Michele Magno}, \bibinfo{person}{Faycal~Ait Aoudia}, \bibinfo{person}{Matthieu Gautier}, \bibinfo{person}{Olivier Berder}, {and} \bibinfo{person}{Luca Benini}.} \bibinfo{year}{2017}\natexlab{}.
\newblock \showarticletitle{WULoRa: An energy efficient IoT end-node for energy harvesting and heterogeneous communication}. In \bibinfo{booktitle}{\emph{Design, Automation \& Test in Europe Conference \& Exhibition (DATE), 2017}}. \bibinfo{pages}{1528--1533}.
\newblock
\urldef\tempurl%
\url{https://doi.org/10.23919/DATE.2017.7927233}
\showDOI{\tempurl}


\bibitem[Magno et~al\mbox{.}(2018)]%
        {magno2018micro}
\bibfield{author}{\bibinfo{person}{Michele Magno}, \bibinfo{person}{Dario Kneub{\"u}hler}, \bibinfo{person}{Philipp Mayer}, {and} \bibinfo{person}{Luca Benini}.} \bibinfo{year}{2018}\natexlab{}.
\newblock \showarticletitle{Micro kinetic energy harvesting for autonomous wearable devices}. In \bibinfo{booktitle}{\emph{2018 International symposium on power electronics, electrical drives, automation and motion (SPEEDAM)}}. IEEE, \bibinfo{pages}{105--110}.
\newblock


\bibitem[Mamatnabiyev(2022)]%
        {zhumani_cattle_tracker_2022}
\bibfield{author}{\bibinfo{person}{Zhumaniyaz Mamatnabiyev}.} \bibinfo{year}{2022}\natexlab{}.
\newblock \showarticletitle{Animal Tracking System Based on GPS Sensor and LPWAN}. In \bibinfo{booktitle}{\emph{2022 International Conference on Smart Information Systems and Technologies (SIST)}}. \bibinfo{pages}{1--4}.
\newblock
\urldef\tempurl%
\url{https://doi.org/10.1109/SIST54437.2022.9945724}
\showDOI{\tempurl}


\bibitem[Mayer et~al\mbox{.}(2020)]%
        {mayer2020smart}
\bibfield{author}{\bibinfo{person}{Philipp Mayer}, \bibinfo{person}{Michele Magno}, {and} \bibinfo{person}{Luca Benini}.} \bibinfo{year}{2020}\natexlab{}.
\newblock \showarticletitle{Smart power unit—mW-to-nW power management and control for self-sustainable IoT devices}.
\newblock \bibinfo{journal}{\emph{IEEE Transactions on Power Electronics}} \bibinfo{volume}{36}, \bibinfo{number}{5} (\bibinfo{year}{2020}), \bibinfo{pages}{5700--5710}.
\newblock


\bibitem[mOOvement(2021)]%
        {MOOVEMENT_gpseartag}
\bibfield{author}{\bibinfo{person}{mOOvement}.} \bibinfo{year}{2021}\natexlab{}.
\newblock \bibinfo{title}{GPS Ear Tag For Cattle}.
\newblock
\newblock
\urldef\tempurl%
\url{https://www.moovement.com.au/gps-ear-tags-for-cattle/}
\showURL{%
\tempurl}
\newblock
\shownote{[Online; accessed 2-June-2023]}.


\bibitem[Munoz~Poblete et~al\mbox{.}(2018)]%
        {munoz_cow_ear_tag_2018}
\bibfield{author}{\bibinfo{person}{Carlos Munoz~Poblete}, \bibinfo{person}{Nicolas Roa~Munoz}, {and} \bibinfo{person}{Juan~Ignacio Huircan~Quilaqueo}.} \bibinfo{year}{2018}\natexlab{}.
\newblock \showarticletitle{Caw’s Walking State Recognition Based on Accelerometers and Gyroscopes Installed on Ear-Tags and Collar-Tags}.
\newblock \bibinfo{journal}{\emph{IEEE Latin America Transactions}} \bibinfo{volume}{16}, \bibinfo{number}{9} (\bibinfo{year}{2018}), \bibinfo{pages}{2490--2495}.
\newblock
\urldef\tempurl%
\url{https://doi.org/10.1109/TLA.2018.8789573}
\showDOI{\tempurl}


\bibitem[Nóbrega et~al\mbox{.}(2018)]%
        {nobrega_2018}
\bibfield{author}{\bibinfo{person}{Luís Nóbrega}, \bibinfo{person}{André Tavares}, \bibinfo{person}{António Cardoso}, {and} \bibinfo{person}{Pedro Gonçalves}.} \bibinfo{year}{2018}\natexlab{}.
\newblock \showarticletitle{Animal monitoring based on IoT technologies}, In \bibinfo{booktitle}{2018 IoT Vertical and Topical Summit on Agriculture - Tuscany (IOT Tuscany)}.
\newblock \bibinfo{journal}{\emph{Nature methods}}, \bibinfo{pages}{1--5}.
\newblock
\urldef\tempurl%
\url{https://doi.org/10.1109/IOT-TUSCANY.2018.8373045}
\showDOI{\tempurl}


\bibitem[Porto et~al\mbox{.}(2021)]%
        {porto_cattle_diseases_detect_2021}
\bibfield{author}{\bibinfo{person}{Simona~M.C. Porto}, \bibinfo{person}{Francesca Valenti}, \bibinfo{person}{Giulia Castagnolo}, {and} \bibinfo{person}{Giovanni Cascone}.} \bibinfo{year}{2021}\natexlab{}.
\newblock \showarticletitle{A Low Power GPS-based device to develop KDE analyses for managing herd in extensive livestock systems}. In \bibinfo{booktitle}{\emph{2021 IEEE International Workshop on Metrology for Agriculture and Forestry (MetroAgriFor)}}. \bibinfo{pages}{243--247}.
\newblock
\urldef\tempurl%
\url{https://doi.org/10.1109/MetroAgriFor52389.2021.9628711}
\showDOI{\tempurl}


\bibitem[Pratama et~al\mbox{.}(2019)]%
        {pratama_cattle_collar_2019}
\bibfield{author}{\bibinfo{person}{Yogi~Putra Pratama}, \bibinfo{person}{Dwi Kurnia~Basuki}, \bibinfo{person}{Sritrusta Sukaridhoto}, \bibinfo{person}{Alviansyah~Arman Yusuf}, \bibinfo{person}{Heri Yulianus}, \bibinfo{person}{Faruq Faruq}, {and} \bibinfo{person}{Fariz~Bramasta Putra}.} \bibinfo{year}{2019}\natexlab{}.
\newblock \showarticletitle{Designing of a Smart Collar for Dairy Cow Behavior Monitoring with Application Monitoring in Microservices and Internet of Things-Based Systems}. In \bibinfo{booktitle}{\emph{2019 International Electronics Symposium (IES)}}. \bibinfo{pages}{527--533}.
\newblock
\urldef\tempurl%
\url{https://doi.org/10.1109/ELECSYM.2019.8901676}
\showDOI{\tempurl}


\bibitem[Rahman et~al\mbox{.}(2018)]%
        {rahman_cattle_behaviour_2017}
\bibfield{author}{\bibinfo{person}{Ashfaqur Rahman}, \bibinfo{person}{Daniel Smith}, \bibinfo{person}{B Little}, \bibinfo{person}{A Ingham}, \bibinfo{person}{Paul Greenwood}, {and} \bibinfo{person}{G.J. Bishop-Hurley}.} \bibinfo{year}{2018}\natexlab{}.
\newblock \showarticletitle{Cattle behaviour classification from collar, halter, and ear tag sensors}.
\newblock \bibinfo{journal}{\emph{Information Processing in Agriculture}}  \bibinfo{volume}{5} (\bibinfo{date}{03} \bibinfo{year}{2018}), \bibinfo{pages}{124--133}.
\newblock
\urldef\tempurl%
\url{https://doi.org/10.1016/j.inpa.2017.10.001}
\showDOI{\tempurl}


\bibitem[Reddy~Maddikunta et~al\mbox{.}(2021)]%
        {reddy_smart_farming_2021}
\bibfield{author}{\bibinfo{person}{Praveen~Kumar Reddy~Maddikunta}, \bibinfo{person}{Saqib Hakak}, \bibinfo{person}{Mamoun Alazab}, \bibinfo{person}{Sweta Bhattacharya}, \bibinfo{person}{Thippa~Reddy Gadekallu}, \bibinfo{person}{Wazir~Zada Khan}, {and} \bibinfo{person}{Quoc-Viet Pham}.} \bibinfo{year}{2021}\natexlab{}.
\newblock \showarticletitle{Unmanned Aerial Vehicles in Smart Agriculture: Applications, Requirements, and Challenges}.
\newblock \bibinfo{journal}{\emph{IEEE Sensors Journal}} \bibinfo{volume}{21}, \bibinfo{number}{16} (\bibinfo{year}{2021}), \bibinfo{pages}{17608--17619}.
\newblock
\urldef\tempurl%
\url{https://doi.org/10.1109/JSEN.2021.3049471}
\showDOI{\tempurl}


\bibitem[Sayeed et~al\mbox{.}(2019)]%
        {sayeed_drig_delivery_2019}
\bibfield{author}{\bibinfo{person}{Md~Abu Sayeed}, \bibinfo{person}{Saraju~P. Mohanty}, \bibinfo{person}{Elias Kougianos}, {and} \bibinfo{person}{Hitten~P. Zaveri}.} \bibinfo{year}{2019}\natexlab{}.
\newblock \showarticletitle{An IoT-based Drug Delivery System for Refractory Epilepsy}. In \bibinfo{booktitle}{\emph{2019 IEEE International Conference on Consumer Electronics (ICCE)}}. \bibinfo{pages}{1--4}.
\newblock
\urldef\tempurl%
\url{https://doi.org/10.1109/ICCE.2019.8661979}
\showDOI{\tempurl}


\bibitem[Schulthess et~al\mbox{.}(2023a)]%
        {schulthess_tinybird_2023}
\bibfield{author}{\bibinfo{person}{Lukas Schulthess}, \bibinfo{person}{Steven Marty}, \bibinfo{person}{Matilde Dirodi}, \bibinfo{person}{Mariana~D. Rocha}, \bibinfo{person}{Linus Rüttimann}, \bibinfo{person}{Richard H.~R. Hahnloser}, {and} \bibinfo{person}{Michele Magno}.} \bibinfo{year}{2023}\natexlab{a}.
\newblock \showarticletitle{TinyBird-ML: An ultra-low Power Smart Sensor Node for Bird Vocalization Analysis and Syllable Classification}. In \bibinfo{booktitle}{\emph{2023 IEEE International Symposium on Circuits and Systems (ISCAS)}}. \bibinfo{pages}{1--5}.
\newblock
\urldef\tempurl%
\url{https://doi.org/10.1109/ISCAS46773.2023.10181431}
\showDOI{\tempurl}


\bibitem[Schulthess et~al\mbox{.}(2023b)]%
        {schulthess_smartcity_2023}
\bibfield{author}{\bibinfo{person}{Lukas Schulthess}, \bibinfo{person}{Tiago Salzmann}, \bibinfo{person}{Christian Vogt}, {and} \bibinfo{person}{Michele Magno}.} \bibinfo{year}{2023}\natexlab{b}.
\newblock \showarticletitle{A LoRa-based Energy-efficient Sensing System for Urban Data Collection}. In \bibinfo{booktitle}{\emph{2023 9th International Workshop on Advances in Sensors and Interfaces (IWASI)}}. \bibinfo{pages}{69--74}.
\newblock
\urldef\tempurl%
\url{https://doi.org/10.1109/IWASI58316.2023.10164426}
\showDOI{\tempurl}


\bibitem[Singhal et~al\mbox{.}(2022)]%
        {singhal_cattle_collar_2022}
\bibfield{author}{\bibinfo{person}{Garima Singhal}, \bibinfo{person}{Priyankar Choudhary}, \bibinfo{person}{Vusirikala Abhishek}, \bibinfo{person}{Seela Sweety}, \bibinfo{person}{Srinivas Subramanian}, {and} \bibinfo{person}{Neeraj Goel}.} \bibinfo{year}{2022}\natexlab{}.
\newblock \showarticletitle{Cattle Collar: An End-to-End Multi-Model Framework for Cattle Monitoring}. In \bibinfo{booktitle}{\emph{2022 IEEE 5th International Conference on Multimedia Information Processing and Retrieval (MIPR)}}. \bibinfo{pages}{401--407}.
\newblock
\urldef\tempurl%
\url{https://doi.org/10.1109/MIPR54900.2022.00079}
\showDOI{\tempurl}


\bibitem[smaxtec(2023)]%
        {SMAXTEC_leistungsumfang}
\bibfield{author}{\bibinfo{person}{smaxtec}.} \bibinfo{year}{2023}\natexlab{}.
\newblock \bibinfo{title}{Halter, animal welfare}.
\newblock
\newblock
\urldef\tempurl%
\url{https://smaxtec.com/de/smaxtec-leistungsumfang/}
\showURL{%
\tempurl}
\newblock
\shownote{[Online; accessed 2-June-2023]}.


\bibitem[Srbinovski et~al\mbox{.}(2015)]%
        {srbinovski2015energy}
\bibfield{author}{\bibinfo{person}{Bruno Srbinovski}, \bibinfo{person}{Michele Magno}, \bibinfo{person}{Brendan O'Flynn}, \bibinfo{person}{Vikram Pakrashi}, {and} \bibinfo{person}{Emanuel Popovici}.} \bibinfo{year}{2015}\natexlab{}.
\newblock \showarticletitle{Energy aware adaptive sampling algorithm for energy harvesting wireless sensor networks}. In \bibinfo{booktitle}{\emph{2015 IEEE Sensors Applications Symposium (SAS)}}. IEEE, \bibinfo{pages}{1--6}.
\newblock


\bibitem[Tran et~al\mbox{.}(2022)]%
        {tran_animal_behaviour_recognition_2022}
\bibfield{author}{\bibinfo{person}{Duc-Nghia Tran}, \bibinfo{person}{Tu~N. Nguyen}, \bibinfo{person}{Phung Cong~Phi Khanh}, {and} \bibinfo{person}{Duc-Tan Tran}.} \bibinfo{year}{2022}\natexlab{}.
\newblock \showarticletitle{An IoT-Based Design Using Accelerometers in Animal Behavior Recognition Systems}.
\newblock \bibinfo{journal}{\emph{IEEE Sensors Journal}} \bibinfo{volume}{22}, \bibinfo{number}{18} (\bibinfo{year}{2022}), \bibinfo{pages}{17515--17528}.
\newblock
\urldef\tempurl%
\url{https://doi.org/10.1109/JSEN.2021.3051194}
\showDOI{\tempurl}


\bibitem[Tsenkov and Tsenev(2017)]%
        {tsenkov_cow_ear_tag_2017}
\bibfield{author}{\bibinfo{person}{Yuriy Tsenkov} {and} \bibinfo{person}{V. Tsenev}.} \bibinfo{year}{2017}\natexlab{}.
\newblock \showarticletitle{Continuous analysis of free-roaming animals' behavior with ear-tag device}. In \bibinfo{booktitle}{\emph{2017 40th International Spring Seminar on Electronics Technology (ISSE)}}. \bibinfo{pages}{1--5}.
\newblock
\urldef\tempurl%
\url{https://doi.org/10.1109/ISSE.2017.8000993}
\showDOI{\tempurl}


\bibitem[uBlox AG(2023)]%
        {noauthor_datasheet_2023}
\bibfield{author}{\bibinfo{person}{uBlox AG}.} \bibinfo{year}{2023}\natexlab{}.
\newblock \bibinfo{title}{Datasheet {GNSS} module ublox {CAM}-{M8}-{FW3}}.
\newblock
\newblock
\urldef\tempurl%
\url{https://content.u-blox.com/sites/default/files/CAM-M8-FW3_DataSheet_%28UBX-15031574%29.pdf}
\showURL{%
\tempurl}
\newblock
\shownote{Accessed: 2023-09-10}.


\bibitem[Veintimilla et~al\mbox{.}(2022)]%
        {veintimilla_cattle_position_2022}
\bibfield{author}{\bibinfo{person}{Javier Veintimilla}, \bibinfo{person}{Mónica Huerta}, {and} \bibinfo{person}{Jose-Ignacio Castillo-Velazquez}.} \bibinfo{year}{2022}\natexlab{}.
\newblock \showarticletitle{Development of System for Monitoring and Geopositioning for Cattle Using IoT}. In \bibinfo{booktitle}{\emph{2022 IEEE ANDESCON}}. \bibinfo{pages}{1--6}.
\newblock
\urldef\tempurl%
\url{https://doi.org/10.1109/ANDESCON56260.2022.9989658}
\showDOI{\tempurl}


\bibitem[Wijers et~al\mbox{.}(2018)]%
        {wijers_lions_2018}
\bibfield{author}{\bibinfo{person}{Matthew Wijers}, \bibinfo{person}{Paul Trethowan}, \bibinfo{person}{Andrew Markham}, \bibinfo{person}{Byron Du~Preez}, \bibinfo{person}{Simon Chamaillé-Jammes}, \bibinfo{person}{Andrew Loveridge}, {and} \bibinfo{person}{David Macdonald}.} \bibinfo{year}{2018}\natexlab{}.
\newblock \showarticletitle{Listening to Lions: Animal-Borne Acoustic Sensors Improve Bio-logger Calibration and Behaviour Classification Performance}.
\newblock \bibinfo{journal}{\emph{Frontiers in Ecology and Evolution}}  \bibinfo{volume}{6} (\bibinfo{date}{10} \bibinfo{year}{2018}), \bibinfo{pages}{171}.
\newblock
\urldef\tempurl%
\url{https://doi.org/10.3389/fevo.2018.00171}
\showDOI{\tempurl}


\end{thebibliography}
\end{document}